\documentclass[12pt]{article}

\title{}

\makeatletter
\long\def\@makemyfntext#1{$^{\rm *}\ $ #1}

\long\def\@myfootnotetext#1{\insert\footins{\footnotesize
    \interlinepenalty\interfootnotelinepenalty 
    \splittopskip\footnotesep
    \splitmaxdepth \dp\strutbox \floatingpenalty \@MM
    \hsize\columnwidth \@parboxrestore
   \edef\@currentlabel{\csname p@footnote\endcsname\@thefnmark}\@makemyfntext
    {\rule{\z@}{\footnotesep}\ignorespaces
      #1\strut}}}

\def\myfootnotetext{\@ifnextchar
     [{\@xfootnotenext}{\xdef\@thefnmark{\thempfn}\@myfootnotetext}}
\makeatother

\newenvironment{numberedlist}
{\begin{list}{\makebox[20pt]{\hss(\arabic{itemno})\enspace}}
             {\usecounter{itemno}\labelwidth 20pt}}{\end{list}}

\newcounter{itemno}

\newcounter{itemno1}

\newcounter{itemno2}

\newcounter{exno}

\newcounter{defno}

\newenvironment{defn}{\refstepcounter{defno}\medskip \noindent {\bf
Definition \thedefno.\ }}{\medskip}

\newcommand{\sep}{\;\vert\;}

\newcommand{\oprove}{\vdash\kern-.6em\lower.7ex\hbox{$\scriptstyle O$}\,}

\newcommand{\Pscr}{{\cal P}}

\newcommand{\pderivation}{{\cal P}\kern -.1em\hbox{\rm -derivation}}
\newcommand{\pderivationl}{{\cal P}\kern -.1em\hbox{\em -derivation}}
\newcommand{\pderivable}{{\cal P}\kern -.1em\hbox{\rm -derivable}}
\newcommand{\pderivablel}{{\cal P}\kern -.1em\hbox{\em -derivable}}
\newcommand{\pderivations}{{\cal P}\kern -.1em\hbox{\rm -derivations}}
\newcommand{\pderivability}{{\cal P}\kern -.1em\hbox{\rm -derivability}}

\newcommand{\all}{\forall}

\newcommand{\ie}{{\em i.e.}}

\newsavebox{\lpartfig}
\newsavebox{\rpartfig}

\newenvironment{exmple}{
 \begingroup \begin{tabbing} \hspace{2em}\= \hspace{3em}\= \hspace{3em}\=
\hspace{3em}\= \hspace{3em}\= \hspace{3em}\= \kill}{
 \end{tabbing}\endgroup}

\newenvironment{example}{
\begingroup  \begin{tabbing} \hspace{2em}\= \hspace{3em}\= \hspace{3em}\=
\hspace{3em}\= \hspace{3em}\= \hspace{3em}\= \hspace{3em}\= \hspace{3em}\= 
\hspace{3em}\= \hspace{3em}\= \hspace{3em}\= \hspace{3em}\= \kill}{
 \end{tabbing} \endgroup }

\newcommand{\lb}{\langle}
\newcommand{\rb}{\rangle}

     {\\* \hspace*{\fill} \end{trivlist}}

\begin{document}
	
\begin{center}
{\Large {\bf 
Incorporating  User Interaction  into Imperative  Languages}}
\\[20pt] 
Keehang Kwon \\
khkwon@dau.ac.kr
\end{center}
	
\noindent {\bf Abstract}: 
In this paper, we  present two new forms of the $write$ statement: one  of the form $write(x);G$
where $G$ is a statement and the other of the form $write(x);D$ where $D$ is a module.
The former is a generalization of traditional $write$ statement and is
quite useful.  The latter is useful for implementing interactive modules.




\section{Introduction}\label{sec:intro}

In this article, we describe a variant of C with some features that are 
inspired by the work of \cite{Jap03}.
These features include

\begin{numberedlist}

\item A $write$ statement dual to the $read$ statement which has the form of
     $write(x); G$. This statement  has the following $new$ semantics: the machine finds
    a value $v$ for $x$ so that the statement $G$  can be successfully completed.

\item A $interactive$ module of the form $write(x); D$ where $D$ is a set of procedure declarations.
This $write$ statement  has the following interchanged semantics:  the $user$ chooses 
    a value $v$ for $x$.

\end{numberedlist}

The notion of interactive methods/modules is quite indispensable in modern imperative languages. 
Interactive  methods interact with the environment, therefore providing some form of 
interactive computing. This paper aims to achieve interaction by providing {\it interactive modules}
 to imperative languages. Thus we allow, within a module, declarations of the form $write(x)$ where $x$ is a variable.
 The intended meaning is that  the value of $x$ is obtained dynamically from the environment.
  To see the usefulness of interactive modules, let us consider the following method which produces the mobile phone number of each employee.

\begin{example}
	write(y); \\
	phone(x) = 		\\
	\>	case Tom : number = 8375;\\
	\>	case Jill : number = 2312;\\
	\>	case Kim : number = y;
\end{example}

In the above, the variable  $y$ - which is Kim's Phone - will be obtained at run time by
requesting the environment to type in Kim's phone number.

Implementing our language poses no serious problem.
 In this paper, we introduce one 
 way of implementing interaction.
Our implementation scheme is the following; when a module is loaded, the variables in the $write$
statements   will be replaced by the input values typed in by the environment.

\section{The Language}\label{sec:logic}

The language is a subset of the core (untyped) C
 with some extensions. It is described
by $G$-, $D$- and $E$-formulas given by the syntax rules below:
\begin{exmple}
\>$G ::=$ \>   $true \sep A \sep x = exp \sep  G;G  \sep write(x);G$ \\  
\>$D ::=$ \>  $ A = G\ \sep \all x\ D \sep D \land D $\\
\>$E ::=$ \>  $ D \sep write(x); E$\\
\end{exmple}
\noindent
 In the above, 
$A$  represents a head of an atomic procedure definition of the form $p(x_1,\ldots,x_n)$. 
A $D$-formula is a set
of procedure declarations. A $E$-formula is an {\it interactive module}.

In the transition system to be considered, a $G$-formula will function as a statement 
and an   $E$-formula  enhanced with the
machine state (a set of variable-value bindings) will constitute  a program.
Thus, a program is a pair   $\lb  E,  \theta\rb$
where  $\theta$ represents the machine state.
$\theta$ is initially  empty  and will be updated dynamically during execution
via the assignment statements. 

 We will  present an interpreter for our language via a proof theory \cite{Khan87,MNPS91}.
 Note that in the initialization phase (denoted by $exec(\Pscr,G,\Pscr')$), our interpreter
 replaces all the variables in $write$ in $\Pscr$
 with new input values from the environment. 
After that, our interpreter proceeds like traditional C interpreter.
 To be specific, it alternates between 
 the  execution phase 
and the backchaining phase.  
In  the  execution phase (denoted by $ex(\Pscr,G,\Pscr')$), it  
executes a statement $G$  with respect to
 $\Pscr$ and
produce a new program $\Pscr'$
by reducing $G$ 
to simpler forms. The rules
(6)-(9) deal with this phase.
If $G$ becomes a procedure call, the machine switches to the backchaining mode. This is encoded in the rule (5). 
In the backchaining mode (denoted by $bc(D,\Pscr,A,\Pscr')$), the interpreter tries 
to find a matching procedure  for a procedure call $A$ inside the module $D$
 by decomposing $D$ into a smaller unit (via rule (4)-(5)) and
 reducing $D$ to  its instance
 (via rule (2)) and then backchaining on the resulting 
definition (via rule (1)).
 To be specific, the rule (2) basically deals with argument passing: it eliminates the universal quantifier $x$ in $\all x D$
by picking a value $t$ for
$x$ so that the resulting instantiation,  $[t/x]D$, matches the procedure call $A$.
 The notation $S$\ seqand\ $R$ denotes the  sequential execution of two tasks. To be precise, it denotes
the following: execute $S$ and execute
$R$ sequentially. It is considered a success if both executions succeed.
Similarly, the notation $S$\ parand\ $R$ denotes the  parallel execution of two tasks. To be precise, it denotes
the following: execute $S$ and execute
$R$  in any order.  It is considered a success if both executions succeed.
The notation $S \leftarrow R$ denotes  reverse implication, \ie, $R \rightarrow S$.

\begin{defn}\label{def:semantics}
Let $G$ be a statement and let $\Pscr$ be a program.
Then the notion of   executing $\lb \Pscr,G \rb$ and producing a new
program $\Pscr'$-- $exec(\Pscr,G,\Pscr')$ --
 is defined as follows:

\begin{numberedlist}

\item    $bc((A = G_1),\Pscr,A,\Pscr_1)\ \leftarrow$  \\
 $ex(\Pscr,G_1,\Pscr_1)$. \% A matching procedure for $A$ is found.

\item    $bc(\all x D,\Pscr,A,\Pscr_1,)\ \leftarrow$  \\
  $bc([t/x]D,\Pscr, A,\Pscr_1)$. \% argument passing

\item    $bc( D_1\land D_2,\Pscr,A,\Pscr_1)\ \leftarrow$  \\
  $bc(D_1,\Pscr, A,\Pscr_1)$. \% look for  a matching procedure in $D_1$.

\item    $bc( D_1\land D_2,\Pscr,A,\Pscr_1)\ \leftarrow$  \\
  $bc(D_2,\Pscr, A,\Pscr_1)$. \% look for a matching procedure in $D_2$

\item    $ex(\lb D,\theta\rb,A,\Pscr_1)\ \leftarrow$     $bc(D,\Pscr, A,\Pscr_1)$. \% $A$ is a procedure call 

\item  $ex(\Pscr,true,\Pscr)$. \% True is always a success.



\item  $ex(\Pscr,x = exp,\Pscr\uplus \{ \lb x,exp' \rb \}) \leftarrow$ $eval(\Pscr,exp,exp')$. \\
 \% In the assignment statement, it evaluates $exp$ to get $exp'$.
  The symbol
$\uplus$ denotes a set union but $\lb x,V\rb$ in $\Pscr$ will be replaced by $\lb x,E' \rb$.

\item  $ex(\Pscr,G_1; G_2,\Pscr_2)\ \leftarrow$ \\
  $ex(\Pscr,G_1,\Pscr_1)$  seqand  $ex(\Pscr_1,G_2,\Pscr_2)$.
\%  a sequential composition 

\item  $ex(\Pscr, write(x); G_1,\Pscr_1)\ \leftarrow$ \\
  choose (and print) a value $v$ for $x$ so that  $ex(\Pscr_1,[v/x]G_1,\Pscr_1)$.
\%  write statement 

\item  $exec(\lb write(x_1)\ldots write(x_n) D, \theta \rb, G,\Pscr_1)\ \leftarrow$  \\
  $read(y_1)\ldots \ldots read(y_n)$  seqand $ex(\lb [y_1,\ldots,y_n]D,\theta \rb, G ,\Pscr_1)$.  \%
  In the initialization phase,  each $x_i$ is replaced with a new input value $y_i$ typed by the
environment.

\end{numberedlist}
\end{defn}

\noindent
If $ex(\Pscr,G,\Pscr_1)$ has no derivation, then the interpreter returns  the failure.
The rule (9)  deals with the new feature.

\section{Examples }\label{sec:modules}

The following  code displays the employee's age to be determined at run time.

\begin{exmple}
 write(y1); \\
 write(y2); \\
 write(y3); \\
 age(x) =  \\
  \>      switch (x) \{ \\
 \>           case tom:  age = y1;   break; \\
  \>          case kim: age = y2;   break;\\
 \>           case sue: age = y3;    break;\\
 \>           default: age = 0;          \\
        \}\\
\end{exmple}

Now consider the procedure call age(tom). The above code will be changed to the one below in the initialization phase assuming
the environment typed in 30,40,22 for y1,y2,y3 respectively.

\begin{exmple}
 age(x) =  \\
\>        switch (x) \{ \\
 \>           case tom:  age = 30;   break; \\
  \>          case kim: age = 40;   break;\\
 \>           case sue: age = 22;    break;\\
 \>           default: age = 0;          \\
        \}\\
\end{exmple}

\noindent Then execution proceeds in the usual way.

\section{Conclusion}\label{sec:conc}

In this paper, we have presented a new form of the $write$ statement of the form $write(x);S$.
This statement is a generalization of traditional write statement 
$write(exp)$, as we can write the latter as $write(x); x == exp$. Here we assume that
 we allow boolean expressions as statements.
In addition, we have presented a notion of  interactive
 modules.  The notion of interactive modules is an indispensable tool in
modern interactive programming.

\section{Acknowledgements}

This work  was supported by Dong-A University Research Fund.

\bibliographystyle{ieicetr}



\end{document}